\documentclass[reqno,11pt, a4paper]{amsart}
\usepackage{amsmath}
\usepackage{amssymb} 
\newtheorem{thm}{Theorem}
\newtheorem{prop}[thm]{Proposition}
\newtheorem{lemma}[thm]{Lemma}

\DeclareMathOperator{\res}{Res}

\def\pa{\partial}
\def\de{\delta}                        
         \def\la{\lambda}

\def\cF{\mathcal{F}}
\def\bla{\bar{\la}}
\def\bt{\bar{t}}        \def\bu{\bar{u}}        \def\bv{\bar{v}}

\renewcommand{\geq}{\geqslant}
\newcommand{\pf}{\noindent{\it Proof \ }}       
\newcommand{\epf}{{$\quad$ \hfill $\Box$}}
\def\eq#1{(\ref{#1})}
\newcommand{\nn}{\nonumber}
\newcommand{\beq}{\begin{equation}}  
\newcommand{\eeq}{\end{equation}}
\def\bea#1\eea{\begin{align}#1\end{align}}   
\def\bes#1\ees{\begin{subequations} \begin{align}#1\end{align}\end{subequations}}   
\def\bss#1\ess{\begin{subequations} #1 \end{subequations}}   
\def\beqa#1\eeqa{\begin{equation} \begin{aligned}#1\end{aligned}\end{equation}}   
\def\f{\frac}
\def\d{\partial}

\def\bB{\bar{B}}

\begin{document}

\title[Hamiltonian structures of 2D Toda reductions]{Reductions of the dispersionless 2D Toda hierarchy and their Hamiltonian structures}
\author[G.Carlet]{Guido Carlet}
\address{CMUC, University of Coimbra, Portugal.}
\email{gcarlet@gmail.com}
\author[P.Lorenzoni]{Paolo Lorenzoni}
\address{Dipartimento di Matematica e Applicazioni, Universit\`a di Milano-Bicocca, Italy.}
\email{paolo.lorenzoni@unimib.it}
\author[A.Raimondo]{Andrea Raimondo}
\address{SISSA - International School for Advanced Studies, Italy.}
\email{andrea.raimondo@sissa.it}
\begin{abstract}
We study finite-dimensional reductions of the dispersionless 2D Toda hierarchy showing that the consistency conditions for such reductions are given by a system of radial Loewner equations. We then construct their Hamiltonian structures, following an approach proposed by Ferapontov.
\end{abstract}
\maketitle

\section{Introduction}
The dispersionless KP and 2D Toda hierarchies are two main examples of hierarchies of equations of hydrodynamic type with an infinite number of dependent variables \cite{TT95}. The problem of the finite-dimensional reduction of these hierarchies consists in finding constraints which are compatible with the flows of the hierarchy and such that the constrained flows are described by a system of equations hydrodynamic type with a finite number of dependent variables.

In the case of the dispersionless KP hierarchy (or Benney hierarchy) it was shown \cite{GT96, GT99} that the equations describing the compatibility of the constraints are a system of chordal Loewner equations. Recently, the Hamiltonian formulation of such reductions have been studied, in terms of nonlocal \cite{GLR} and purely nonlocal \cite{GLR2} Poisson brackets.

In this paper we consider $N$-dimensional reductions of the dispersionless 2D Toda hierarchy. In Section 3 we show that the consistency of a reduction is equivalent to a system of radial Loewner equations. We show that the compatibility conditions for such system is given by Gibbons-Tsarev equations from which it follows that the reductions are semi-Hamiltonian. We present generating functions for the flows (or symmetries) of the reductions and give a proof of the functional dependence of the two Lax functions. In Section 4 we consider the Hamiltonian structures of the reductions. Following an approach of Ferapontov we show that it is possible to factorize the Riemann curvature tensor, associated with some diagonal metric, in terms of the symmetries of the reduction. From such factorization the existence of nonlocal Hamiltonian structures follows. In Section 5 purely nonlocal Hamiltonian structures for the reductions are studied and in Section 6 an example is considered.

\section{The dispersionless 2D Toda hierarchy}

The Lax representation of the dispersionless two-dimensional Toda hierarchy \cite{TT95} is defined in terms of two formal Laurent power series in $p$
\begin{subequations} \label{hi1+2}
\begin{align} \label{hi1}
&\la = p + u_0 + u_1 p^{-1} + \dots \\
&\bla = \bu_{-1} p^{-1} +  \bu_0 + \bu_1 p + \dots \label{hi2} .
\end{align}
\end{subequations}
where the dependent variables $u_k$ and $\bar{u}_k$ depend on the spatial variable $x$ and on two infinite sets of independent variables $t_n$ and $\bt_n$ for $n >0$. The Lax equations are
\begin{subequations} \label{hi4+5}
\begin{align} \label{hi4}
&\la_{t_n} = \{ B_n , \la \}, \qquad \bla_{t_n} = \{ B_n , \bla \} , \\
&\la_{\bt_n} = \{ \bB_n , \la \}, \qquad \bla_{\bt_n} = \{ \bB_n , \bla \} ,\label{hi5}
\end{align}
\end{subequations}
where the Poisson brackets are given by
$$\{ f, g \} = p \frac{\pa f}{\pa p} \frac{\pa g}{\pa x} - p \frac{\pa g}{\pa p} \frac{\pa f}{\pa x},$$
and we define
\begin{equation*}
B_n := \frac1n (\la^n)_+ \qquad \bB_n := \frac1n (\bla^n)_-  .
\end{equation*}
We denote by $(\ )_+$ and $(\ )_-$ the projections of a power series in $p$ to positive and strictly negative powers of $p$ respectively. In this formal setting the Lax equations are considered as generating functions of an infinite set of involutive evolutionary equations of hydrodynamic type for the coefficients $u_k$ and $\bu_k$. 




In the following we will consider $\la$ and $\bla$ as univalent analytic functions on certain domains in the complex plane, having the expansions \eq{hi1+2} at $p=\infty,0$ respectively. 

\section{Reductions of the 2D Toda hierarchy}
In this section we consider the reductions of the dispersionless 2D Toda hierarchy and their relation with systems of radial Loewner equations. Similar results were first obtained for the dispersionless KP case by Gibbons and Tsarev \cite{GT96,GT99}; other examples -- including the dispersionless $2D$ Toda hierarchy -- have been studied, for instance, in \cite{yu,GMA03,TTZ,TT06,TT08}.

A \emph{reduction} of the dispersionless 2D Toda hierarchy is given by a choice of two families of functions 
\begin{equation} \label{re1}
\la = \la(p; \la_1, \dots , \la_N), \qquad \bla = \bla( p; \la_1, \dots , \la_N), 
\end{equation}
on the $p$-plane depending on $N$ parameters $\la_1, \dots , \la_N$ such that the flows \eq{hi4+5} are consistent with \eq{re1} and are induced by diagonal hydrodynamic type equations 
\beq \label{diag}
\frac{\pa \la_i}{\pa t_n} = v^i_{(n)} \frac{\pa \la_i}{\pa x}, \quad 
\frac{\pa \la_i}{\pa \bt_n} = \bv^i_{(n)}  \frac{\pa \la_i}{\pa x},
\eeq
where $v^i_{(n)}$ and $\bv^i_{(n)}$ are suitable functions of $\la_1, \dots , \la_N$, depending on the choice of $\la$ and $\bla$. We will denote by $v^i=v^i_{(1)}$ the first of these functions.
\begin{prop}
The functions \eq{re1} with \eq{diag} provide a reduction of the dispersionless 2D Toda hierarchy if and only if the following system of Loewner equations is satisfied
\beq \label{loewner}
\frac{\pa \la}{\pa \la_i} = \frac{p \la_p}{p-v^i} \frac{\pa u_0}{\pa \la_i} , \quad
\frac{\pa \bla}{\pa \la_i} = \frac{p \bla_p}{p-v^i} \frac{\pa u_0}{\pa \la_i}.
\eeq
Moreover, the functions $v^i_{(n)}$ and $\bv^i_{(n)}$ appearing in \eqref{diag} are given by
\beq \label{vien}
v^i_{(n)} = \left(p \frac{\pa B_n}{\pa p}\right)_{|_{p=v^i}} , \quad
\bv^i_{(n)} = \left(p \frac{\pa \bB_n}{\pa p}\right)_{|_{p=v^i}} .
\eeq
\end{prop}
\pf Assuming the independence of the $\la^i_x$, the Lax equations \eqref{hi4} -- together with conditions \eq{re1} and \eq{diag} -- are equivalent to the following set of equations:
\beq \label{re3}
\frac{\pa \la}{\pa \la_i} = \frac{p \la_p }{p (B_n)_p - v^i_{(n)}} \frac{\pa B_n}{\pa \la_i}, \quad
\frac{\pa \bla}{\pa \la_i} = \frac{p \bla_p}{p (B_n)_p - v^i_{(n)}}  \frac{\pa B_n}{\pa \la_i} . 
\eeq
Conditions for the flow $\bt_n$ are similar
\beq \label{re7}
\frac{\pa \la}{\pa \la_i} = \frac{p \la_p}{p (\bB_n)_p - \bv^i_{(n)}} \frac{\pa \bB_n}{\pa \la_i}, \quad
\frac{\pa \bla}{\pa \la_i} = \frac{p \bla_p}{p (\bB_n)_p - \bv^i_{(n)}} \frac{\pa \bB_n}{\pa \la_i} .
\eeq
For $n=1$ we have $B_1 = p + u_0$; therefore, in this case equations \eq{re3} are exactly the Loewner system \eq{loewner}. We prove now that the other conditions \eq{re3} -- for $n\geq2$ -- and \eq{re7}, follow from \eq{loewner}. To see this, consider the first equation in \eq{re3} for $n \geq2$. Using Loewner equations, we have
\beq
\frac{\pa B_n}{\pa \la_i} = \frac1n \frac{\pa (\la^n)_+}{\pa \la_i} 
= \frac1n \left( \frac{p(\la^n)_p}{p-v^i} \right)_+ \frac{\pa u_0}{\pa \la_i} .\notag
\eeq
On the other hand, using the fact that $\frac1{p-v^i} = \left(\frac1{p-v^i}\right)_-$, we see that
\bea
\frac1n \left( \frac{p (\la^n)_p }{p-v^i} \right)_+ &= \frac1n\left( \frac{(p (\la^n)_p)_+ - ((p (\la^n)_p)_+)|_{p=v^i} }{p-v^i} \right)_+\notag\\
&=\frac{p(B_n)_p - v^i_{(n)}}{p-v^i}.\notag
\eea
Using these formulas we obtain
\bea
\frac{p \la_p }{p (B_n)_p - v^i_{(n)}} \frac{\pa B_n}{\pa \la_i} = \frac{p \la_p}{p-v^i} \frac{\pa u_0}{\pa \la_i}\notag
\eea
hence the first equation in \eq{re3} is a consequence of the Loewner equations \eq{loewner}. Analogously, all other conditions \eq{re3} and \eq{re7} are proved.
\epf
\newline

The Loewner equations can be equivalently written in terms of $\varphi := \log \bu_{-1}$ instead of $u_0$. Indeed, expanding both sides of the Loewner equations \eq{loewner} at $p\sim \infty$, we get
\beq \label{u0phi}
\frac{\pa u_0}{\pa \la_i}= v^i \frac{\pa \varphi}{\pa \la_i},
\eeq
and substituting back into the Loewner system, we obtain
\beq \label{loewnerphi}
\frac{\pa \la}{\pa \la_i} = \frac{v^i p \la_p}{p-v^i} \frac{\pa \varphi}{\pa \la_i} , \qquad
\frac{\pa \bla}{\pa \la_i} = \frac{v^i p \bla_p}{p-v^i} \frac{\pa \varphi}{\pa \la_i}.
\eeq
We derive now the compatibility conditions of the above system.

\begin{prop}
The Loewner equations \eqref{loewnerphi} are compatible if and only if the functions $v^i$ and $\varphi$ satisfy the Gibbons-Tsarev equations
\bss \label{GT}
\bea \label{GT1}
\frac{\pa v^i}{\pa \la_j} &= \frac{v^i\, v^j}{v^j-v^i} \frac{\pa \varphi}{\pa \la_j},&i\neq j,\\
\frac{\pa^2 \varphi}{\pa \la_i \pa \la_j} &= 2\, \frac{v^i\, v^j}{(v^i-v^j)^2} \frac{\pa \varphi}{\pa \la_i} \frac{\pa \varphi}{\pa \la_j},&i\neq j. \label{GT2}
\eea
\ess
\end{prop}
\pf
Spelling out the compatibility conditions
\beq \label{comp}
\frac{\pa}{\pa \la_i} \frac{\pa}{\pa \la_j} \la = \frac{\pa}{\pa \la_j} \frac{\pa}{\pa \la_i} \la,
\eeq
we obtain
\bea
&\frac{\pa_i u_0 \pa_j v^i}{(p-v^i)^2} - \frac{\pa_j u_0 \pa_i v^j}{(p-v^i)^2} + \pa_i \pa_j u_0 \left( \frac1{p-v^i} - \frac1{p-v^j} \right) +\notag\\
&+ p \frac{\pa_i u_0 \pa_j u_0}{(p-v^i)(p-v^j)}\left( \frac1{p-v^i} - \frac1{p-v^j} \right) =0 ,\notag
\eea
where we have used the notation $\pa_i = \frac{\pa}{\pa \la_i}$.
The Gibbons-Tsarev equations can be recovered as the two leading coefficients of the expansion for $p \sim v^i$. It is easily checked by direct substitution in \eq{comp} that the Gibbons-Tsarev are also sufficient for the compatibility.
\epf
\newline

Using the Gibbons-Tsarev system \eq{GT}, it is easy to prove that the characteristic velocities $v^i$ of the reduction of the dispersionless 2D Toda hierarchy satisfy the condition
\beq \label{semiham}
\frac{\pa}{\pa \la_k} \left( \frac{\frac{\pa v^i}{\pa \la_j}}{v^i-v^j} \right) =
\frac{\pa}{\pa \la_j} \left( \frac{\frac{\pa v^i}{\pa \la_k}}{v^i-v^k} \right),\qquad i\neq j\neq k\neq i,
\eeq
known in the literature as semi-Hamiltonian condition \cite{T90}. The above system arises as the compatibility condition of the linear system
\beq\label{sym}
\frac{\pa w^i}{\pa \la_j}= \frac{\frac{\pa v^i}{\pa \la_j}}{v^i-v^j}\left(w^i-w^j\right), \qquad i\neq j,
\eeq
which gives the characteristic velocities $w^i$ of the symmetries of the reduction. Therefore, every $N$-tuple of functions defined in \eq{vien} automatically satisfies the semi-Hamiltonian condition \eq{semiham}.


\smallskip
Next, we define a pair of generating functions for the symmetries.

\begin{lemma}
The functions defined by
\beq \label{defW}
W^i (\la)= \frac{p(\la)}{p(\la)-v^i}, \qquad
\bar{W}^i (\bla) = \frac{p(\bla)}{p(\bla)-v^i}
\eeq
where $p(\la):=\la(p)^{-1}$ and $p(\bla):=\bla(p)^{-1}$ are the inverse functions of $\la(p)$ and $\bla(p)$ respectively, are generating functions for the characteristic velocities \eq{vien}, namely
\beq
W^i(\la) = \sum_{n\geq0} v^i_{(n)} \la^{-n}, \qquad \bar{W}^i(\bla) = \sum_{n\geq1} \bv^i_{(n)} \bla^{-n}.\notag
\eeq
\end{lemma}
\pf
The function $W^i$ has an asymptotic expansion for $\la\mapsto\infty$ of the form
\beq
W^i(\la) = \sum_{n\geq0} c^i_{(n)} \la^{-n}. \notag
\eeq
We want to prove that $c^i_{(n)}=v^i_{(n)}$. Clearly, the coefficient of $\la^{-n}$ in the expansion of $W^i$ is given by
$$c^i_{(n)}=-\underset{\la = \infty}{\res}\, W^i(\la) \la^{n-1} d\la.$$
Expressing the residue in the variable $p$ we rewrite it as
$$- \frac1n\,\,\underset{p=\infty}{\res}\,\,\frac p{p-v^i}\, (\la^n)_p\, dp=-\frac{v^i}{n}\,\,\underset{p=\infty}{\res}\,\,\frac{(\la^n)_+}{(p-v^i)^2}dp$$
where in the right hand side we have integrated per parts and dropped the irrelevant negative powers of $p$ in the numerator. We are left with the residue of a rational expression with poles only at $p=\infty,v^i$. Hence it is equal to
$$c^i_{(n)}=\frac{v^i}n\,\,\underset{p=v^i}{\res}\,\,\frac{(\la^n)_+}{(p-v^i)^2} dp = \frac{v^i}{n} (((\la^n)_+)_p)_{|_{p=v^i}} = v_{(n)}^i .$$
An analogous proof holds for the generating function $\bar{W}^i$.
\epf
\newline


In the next section, we will find convenient to use different generating functions, which are obtained (up to a sign) by differentiating $W^i$ and $\bar{W}^i$ with respect to $\la$:
\bes
&w^i(\la) = \frac{v^i}{(p(\la)-v^i)^2} \frac{\pa p(\la)}{\pa \la} = \sum_{n\geq1} n v^i_{(n)} \la^{-n-1}, \label{gsym1} \\
&\bar{w}^i(\bla) = \frac{v^i}{(p(\bla)-v^i)^2} \frac{\pa p(\bla)}{\pa \bla} = \sum_{n\geq1} n \bv^i_{(n)} \bla^{-n-1}.
\ees
\newline
In the reductions of the dispersionless $2D$ Toda hierarchy one needs only one of the two Lax functions $\la$, $\bla$; indeed, we have the following
\begin{prop} For a reduction of the dispersionless $2D$ Toda hierarchy,  the univalent functions $\la$ and $\bar{\la}$ are functionally dependent on the common domain of definition.
%
\end{prop}
\pf
On any domain of the complex plane where $\bla$ is invertible, denote $p(\bla,\la_1, \dots, \la_N )$ its inverse and define
$$\cF(\bla) =\la(p(\bla,\la_1,\dots,\la_N),\la_1,\dots,\la_N),$$
which is well-defined on the image by $\bla$ of the intersection of the domains of definition of $\la$ and $\bar{\la}$. A priori $\cF$ might depend on  $\la_1, \dots,\la_N$, however since $\la$ and $\bla$ satisfy the Loewner equations \eq{loewner} we have
$$\frac{\pa \la}{\pa \la_i} = \frac{\pa \cF}{\pa \bla} \frac{\pa \bla}{\pa \la_i} + \frac{\pa \cF}{\pa \la_i} 
= \frac{\pa \cF}{\pa \bla} \frac{p\bla_p}{p-v^i}\frac{\pa u_0}{\pa \la_i} + \frac{\pa \cF}{\pa \la_i} 
= \frac{\pa \la}{\pa \la_i} + \frac{\pa \cF}{\pa \la_i}$$
hence $\frac{\pa \cF}{\pa \la_i}=0$. This shows that $\la$ can be expressed in terms of $\bla$ in terms of the function $\cF$ which is independent of the parameters $\la_1, \dots, \la_N$. 
\epf
\newline

In particular, it follows from the previous Proposition that $\la$ and $\bla$ have the same critical points. Moreover, we have a relation between the critical points of $\la$ and the characteristic velocities of the reduction. Indeed, evaluating the $2D$ Toda equations at a critical point, one easily proves that
\begin{prop}
If $\hat{p}=\hat{p}(\la_1, \dots, \la_N )$ is a critical point of $\la$, i.e. $\la_p(\hat{p})=0$, and $\hat\la := \la(\hat{p};\la_1,\dots ,\la_N)$ the corresponding critical value then 
$$\frac{\pa \hat\la}{\pa t_n} = v_{(n)} \frac{\pa \hat\la}{\pa x}, \quad \frac{\pa \hat\la}{\pa \bt_n} = \bar{v}_{(n)}  \frac{\pa \hat\la}{\pa x},$$
with 
$$v_{(n)} = \left(p \frac{\pa B_n}{\pa p}\right)_{|_{p=\hat{p}}} , \quad
\bar{v}_{(n)} = \left(p \frac{\pa \bB_n}{\pa p}\right)_{|_{p=\hat{p}}}.$$
\end{prop}
On the other hand, under generic assumptions, the characteristic velocities are critical points of $\la$ (and $\bla$):
\begin{prop}
The characteristic velocities $v^j$ are critical points of $\la$ and the Riemann invariants can be chosen to be the corresponding critical values.
\end{prop}
\pf
From the Loewner equation evaluated at $p=v^j$ one has
$$\frac{\pa \la}{\pa \la_i}_{|p=v^j} (v^j-v^i) = v^j \la_p(v^j) \frac{\pa u_0}{\pa \la_i},$$
which for $i=j$ implies $\la_p(v^j)=0$. 

\epf
\newline

In the rest of the article we will assume that 
the Riemann invariants $\la_i$ are the critical values corresponding to the critical points $v^i$. Under this assumption, we have the following
\begin{lemma}\label{ndderprop}
The formula
$$\frac{\partial\varphi}{\partial\lambda_i}=\frac{1}{\left(v^i\right)^2\lambda_{pp}(v^i)}$$
holds for any reduction of the Toda hierarchy.
\end{lemma}
\pf
Considering the Loewner equation \eqref{loewner} and taking the limit for $p\to v^i$ one gets
$$1=\lim_{p\to v^i}\frac{p\lambda_p(p)}{p-v^i}\frac{\partial u_0}{\partial\lambda_i}=\lim_{p\to v^i}\frac{p\left(\lambda_p(p)-\lambda_p(v^i)\right)}{p-v^i}\frac{\partial u_0}{\partial\lambda_i}= v^i\lambda_{pp}(v^i)\frac{\partial u_0}{\partial\lambda_i},$$
which holds since $v^i$ is a critical point of $\la$. The thesis follows from identity \eqref{u0phi}.
\epf
\newline

\section{Hamiltonian formulation}
We have seen that the reductions of the dispersionless $2D$ Toda hierarchy are semi-Hamiltonian systems of hydrodynamic type.  In \cite{Fera1} Ferapontov conjectured that
 any semi-Hamiltonian system is always Hamiltonian with respect to suitable,
 possibly nonlocal, Hamiltonian operators (see also \cite{F95}), which are obtained by the following construction:
\smallskip

1. Find the general solution  of the system
\begin{equation}
\label{meq}
\frac{ \pa }{\pa \la_j}\ln{\sqrt{g_{ii}}}=\f{\frac{ \pa v^i}{\pa \la_j}}{v^j-v^i},\qquad i\neq j.
\end{equation}
To this purpose is sufficient to find one solution $g_{ii}$  of (\ref{meq}), since 
the general solution is  $\f{g_{ii}}{\phi^i(\la_i)}$, where $\phi^i$ are arbitrary functions
 of one argument. The functions $g^{ii}$ define the non-vanishing controvariant components of
 a diagonal metric.
\smallskip

2. Write the non-vanishing components of the curvature tensor in terms
 of solutions $w_{\alpha}^i$ of the linear system (\ref{sym}):
\begin{equation}\label{exp}
R^{ij}_{ij}=\sum_{\alpha}\epsilon_\alpha w^i_{\alpha}w^j_{\alpha}\hspace{1 cm}\epsilon_{\alpha}=\pm 1.
\end{equation}
Given a solution of (\ref{meq}) and the quadratic expansion \eq{exp} of the associated curvature tensor,
 the Hamiltonian structure is given by
\begin{equation*}
\Pi^{ij}=g^{ii}\delta^{ij}\frac d{dx}+\Gamma^{ij}_k(\la) \la_{k,x}
+\sum_{\alpha}\epsilon_\alpha w^i_{\alpha}\la_{i,x}\left(\frac{d}{d
x}\right)^{-1}\!\!w^j_{\alpha}\la_{j,x},
\end{equation*}
where $\Gamma^{ij}_k=-g^{ii}\Gamma^{j}_{ik}$ and the $\Gamma^{j}_{ik}$ are the Christoffel symbols of the metric $g$. We recall that the index $\alpha$ can take values on a finite or infinite - even continuous - set.
\smallskip

We now apply the above procedure to find a Hamiltonian formulation for reductions of the dispersionless $2D$ Toda hierarchy.
Using the Gibbons-Tsarev equations it is easy to check that
\begin{prop}
The general solution of the system \eq{meq} is given by
\begin{equation} \label{me2}
g_{ii} = \frac1{\phi_i (\lambda_i)}\frac{ \pa \varphi}{\pa \la_i},
\end{equation}
where each $\phi_i$ is a function of the sole variable $\lambda_i$.
\end{prop}
\smallskip

Let us consider first the case of potential metric, given by
\begin{equation} \label{me3}
g_{ij} =  \frac{\pa\varphi}{\pa\la_i}   \de_{ij} .
\end{equation}


Following the Ferapontov's procedure we have now to find a quadratic expansion of the form \eq{exp} for the curvature tensor of the metric \eq{me3}. For this purpose, it is convenient to introduce the following function
\beq \label{ri0}
F(p(\la);\lambda_1,\dots,\lambda_N) = \frac{\pa p}{\pa \la} + \sum_{j=1}^N \frac{\pa p}{\pa \la_j},\notag
\eeq
which can be expressed in terms of the variable $p$ as
\beq\label{fp}
F(p;\lambda_1,\dots,\lambda_N)=\frac1{\la_p} - \sum_{j=1}^N\f{p\,  v^j}{p-v^j} \frac{\pa \varphi}{\pa \la_j}.
\eeq
Here we used the identity
\beq\label{loewnerp}
\frac{\pa p}{\pa \la_i} = \frac{p\,v^i}{v^i-p} \frac{\pa \varphi}{\pa \la_i},
\eeq
which follows from the Loewner equations \eq{loewnerphi}. We need the following technical lemma:
\begin{lemma}\label{techlem}
The function $F$ is analytic at $p=v^i$, $i=1 , \dots, N$ and satisfies
\bea
F_{|p=v^i} &=\delta(v^i) \label{ri3}, \\
\f{\d F}{\d p}_{|p=v^i} &= \delta \left(\log \sqrt{\frac{\pa\varphi}{\pa\la_i}}\right) +  \frac{\de (v^i)}{v^i},  \label{ri5} 
\eea
where the operator $\de$ is given by
$$\de := \sum_{k=1}^N \frac{\pa}{\pa\la_k}.$$
\end{lemma}
\pf
Using Lemma \ref{ndderprop}  it is easy to see that the poles of $\frac1{\la_p}$ in $v^i$ cancel out in \eq{fp}, hence $F$ is analytic at $p=v^i$. Moreover, if we consider the compatibility condition between the Loewner system \eq{loewnerp} and the equation
$$\frac{\pa p}{\pa \la} = F + \sum_{j=1}^N \frac{p \, v^j}{p-v^j} \frac{\pa \varphi}{\pa \la_j},$$
we obtain
\begin{align}\label{ri9}
0=\frac{\pa^2 p}{\pa \la_i \pa \la} - \frac{\pa^2 p}{\pa \la \pa \la_i} &=- \frac{(v^i)^2}{(p-v^i)^2}   \frac{\pa \varphi}{\pa \la_i}  \left[  F(p) -\de(v^i) \right]\notag \\
&+\frac{(v^i)^2}{p-v^i}   \frac{\pa \varphi}{\pa \la_i}   \left[ -F_p(p) + \delta ( \log \frac{\pa \varphi}{\pa \la_i} ) + 2 \frac{\de( v^i)}{v^i} \right] \\
&+ \text{regular function at $p=v^i$}.\notag
\end{align}
Multiplying by $(p-v^i)^2$ and taking the limit for $p\to v^i$ we get
$$(v^i)^2 \frac{\pa \varphi}{\pa \la_i} ( F(v^i) - \delta(v^i) ) = 0,$$
which, under the assumptions $\frac{\pa\varphi}{\pa \la_i} \ne 0$, $v^i\ne0$, implies identity \eq{ri3}. Computing the residue of \eq{ri9} at $p=v^i$ one obtains \eq{ri5}.
\epf
\newline

We are now in the position to find a quadratic expansion for the curvature of the potential metric \eq{me3}. Indeed, let $\Gamma_i$ be a small contour surrounding the  point $p=v^i$ counter-clockwise and let $C_i$ be the image of $\Gamma_i$ under the map $\lambda$. Let $\Gamma := \bigcup_{i=1}^N \Gamma_i$ and $C:= \bigcup_{i=1}^N C_i$.

\begin{thm}
The non-vanishing components of the Riemann tensor of the potential metric \eq{me3} admit the following quadratic expansion
\begin{equation}\label{mainid3}
R^{ij}_{ij}=-\frac{1}{2\pi i}\int_{C} w^i(\lambda)
w^j(\lambda)d\lambda, \qquad i \not= j,
\end{equation}
where the $w^i(\la)$ are the generating functions of the symmetries defined in \eq{gsym1}.
\end{thm}
\pf
In order to determine the Riemann curvature tensor for the metric \eq{me3}, we use the following well-known fact: {\it
The only non-zero components of the curvature tensor of a diagonal metric $g_{ij}=\de_{ij} g_{ii}$ having symmetric rotation coefficients $\beta_{ij} := \frac{\pa_i \sqrt{ g_{jj}}}{\sqrt{g_{ii}}}$ are:
\begin{equation} \label{cu3}
R_{ijij} =-\sqrt{g_{ii}} \sqrt{g_{jj}} \de ( \beta_{ij} ),\qquad i\neq j.
\end{equation}
}
Using the Gibbons-Tsarev equations we find that in our case the rotation coefficients are given by:
\begin{equation*}
\beta_{ij} = \frac12 \frac{\pa_i \pa_j \varphi}{ \sqrt{\pa_i \varphi} \sqrt{ \pa_j \varphi}} = \sqrt{\pa_i \varphi} \sqrt{\pa_j \varphi} \frac{v^i v^j}{(v^i - v^j)^2},\qquad i\neq j.
\end{equation*}
Substituting into \eq{cu3}, and raising the first two indices, we obtain the formula
\begin{equation} \label{cu2} 
R^{ij}_{ij}=-\f{(v^i+v^j)[v^i \delta(v^j)- v^j \delta(v^i)]}{(v^i-v^j)^3}-
\f{v^i v^j[\delta(\log{\sqrt{\d_i
\varphi }})+\delta(\log{\sqrt{\d_j
\varphi}})]}{(v^i-v^j)^2}, \quad i\not= j,
\end{equation}
which holds for any reduction. We can now use Lemma \ref{techlem} to write \eq{cu2} in the form
\begin{equation*}
R^{ij}_{ij}=-\frac{v^i v^j}{(v^i-v^j)^2} \left(
\frac{\partial F}{\partial p}(v^i)+ \frac{\partial F}{\partial p}(v^j)\right)
+\frac{2 v^i v^j}{(v^i-v^j)^3}\left( F(v^i)-F(v^j)\right) .
\end{equation*}
Moreover, using the fact that $F(p)$ is regular at all $p=v^k$ one can rewrite this expression in terms of residues, obtaining
\begin{equation*}
R^{ij}_{ij} =-\frac1{2 \pi i} \int_{\Gamma} \frac{F(p)v^iv^j}{(p-v^i)^2(p-v^j)^2}\,dp  , \qquad i\not= j .
\end{equation*}
Due to \eq{fp}, the above integral splits in two
$$\frac1{2 \pi i} \int_{\Gamma} \frac{F(p)v^iv^j}{(p-v^i)^2(p-v^j)^2}\,dp=\frac{v^i v^j}{2 \pi i} \int_{\Gamma} \frac{\frac1{\la'(p)}}{(p-v^i)^2 (p-v^j)^2} dp-\frac{v^i v^j}{2 \pi i} \int_{\Gamma} \frac{\sum_{k=1}^N \frac{p v^k \pa_k \varphi}{p-v^k}}{(p-v^i)^2 (p-v^j)^2} dp,$$
and the second term of the right hand side above is zero, for all poles of the rational integrand lie inside the contour $\Gamma$. Hence, we have
\begin{equation*}
R^{ij}_{ij}=-\frac{v^i v^j}{2 \pi i} \int_{\Gamma} \frac{\frac1{\la'(p)}}{(p-v^i)^2 (p-v^j)^2} dp,
\end{equation*}
after a change of the variable of integration, we get 
\begin{equation*}
R^{ij}_{ij}=-\frac{1}{2 \pi i} \int_{C} \frac{v^i\,\frac{\partial p}{\partial \la}}{(p(\la)-v^i)^2}\frac{v^j\,\frac{\partial p}{\partial \la}}{(p(\la)-v^j)^2} d\la,
\end{equation*}
which is exactly formula \eq{mainid3}.
\epf
\newline

%
%

We can now formulate our main theorem on the Hamiltonian representation of the hierarchy in the case of potential metric.
\begin{thm} 
The reduction of the Toda hierarchy associated with the function $\la(p,\la_1,\dots,\la_N)$
is Hamiltonian with the Hamiltonian structure 
\begin{align*}
\Pi^{ij}=&\frac{1}{\partial_i \varphi}\delta^{ij}\frac{d}{dx}+\Gamma^{ij}_k\,\la^k_x\notag\\
&-\frac{1}{2\pi i}\int_{C}\frac{v^i\,\,\frac{\partial
p}{\partial\la}\,\,\,\la^i_x}{(p(\la)-v^i)^2}\left(\frac{d}{d
x}\right)^{-1}\!\!\!\frac{v^j\,\,\frac{\partial
p}{\partial\la}\,\,\,\la^j_x}{(p(\la)-v^j)^2}\,d\la.
\end{align*}
Here
\begin{equation*}
\Gamma^{ij}_k = -\frac1{2 \pa_i \varphi \, \pa_j \varphi} \left( \de_{ij} \pa_i \pa_k \varphi + \de_{jk} \pa_i \pa_j \varphi - \de_{ik} \pa_i \pa_j \varphi\right)
\end{equation*}
are the Christoffel symbols of the metric.
\end{thm}

%
%

%
%

In the general case, with the non-potential metric
\beq\label{genmet}
\left(g_\phi\right)_{ii}=\frac{1}{\phi_i(\la_i)}\frac{\partial \varphi}{\partial \lambda_i}, 
\eeq
we can prove that we have a similar expansion of the curvature tensor, given by
\begin{equation*}
R^{ij}_{ij}=
-\frac{v^i\,v^j}{2\pi i}\sum_{k=1}^N\int_{\Gamma_k}
\frac{\frac{1}{\la_p}
}{(p-v^i)^2(p-v^j)^2}\,\,\phi_k(\la(p))\, dp. 
\end{equation*}
Therefore, we have the following family of Hamiltonian structures
\begin{align*}
\Pi^{ij}=&\phi_i\frac{1}{\partial_i \varphi}\delta^{ij}\frac{d}{dx}+\Gamma^{ij}_k\,\la^k_x\notag\\
&-\frac{1}{2\pi i}\sum_{k=1}^N\int_{C_k}\frac{v^i\,\,\frac{\partial
p}{\partial\la}\,\,\,\la^i_x}{(p(\la)-v^i)^2}\left(\frac{d}{d
x}\right)^{-1}\!\!\!\frac{v^j\,\,\frac{\partial
p}{\partial\la}\,\,\,\la^j_x}{(p(\la)-v^j)^2}\,\phi_k(\la)\,d\la,
\end{align*}
for any choice of the functions $\phi_i$.

\section{Purely nonlocal Hamiltonian structures}
In addition to the nonlocal Ferapontov-type Hamiltonian operators, we can associate to any reduction of the dispersionless $2D$ Toda hierarchy a family of purely nonlocal Hamiltonian operators. In the semi-Hamiltonian case, it has been shown in \cite{GLR2} that if $W^i_\alpha$ are the characteristic velocities of pairwise commuting diagonal hydrodynamic flows, the operator 
$$
\Pi^{ij}=\sum_{\alpha}\epsilon_\alpha W^i_{\alpha}\la^i_x\left(\frac{d}{d
x}\right)^{-1}\!\!\!W^j_{\alpha}\la^j_x,
$$
defines a purely nonlocal Hamiltonian structure provided
\beq
\sum_{\alpha}\epsilon_\alpha W^i_{\alpha}W^j_{\alpha} = 0, \quad i \not= j. \nn
\eeq
Moreover 
$$g^{ii}\delta^{ij}=\sum_{\alpha}\epsilon_\alpha W^i_{\alpha}W^j_{\alpha}.$$
defines a solution to \eq{meq}.

For the reductions of the dispersionless $2D$ Toda hierarchy, the following result holds:
\begin{lemma}
The contravariant components of the metric \eqref{genmet} admit the following quadratic expansion 
\beq\label{qem}
g_\phi^{ii}\,\delta^{ij}=\phi_i(\la_i)\frac{1}{\partial_i\varphi}\delta^{\,ij}=\frac{1}{2\pi  i}\sum_{k=1}^N\int_{C_k}W^i(\lambda)W^j(\lambda)\phi_k(\lambda)\,d\lambda,\notag
\eeq
where the $W^i(\lambda)$ are the generating functions of the symmetries \eqref{defW}.
\end{lemma}
\pf
The proof is a straightforward computation of the integral:
\begin{gather*}
\frac{1}{2\pi i}\sum_{k=1}^N\int_{C_k}W^i(\lambda)W^j(\lambda)\phi_k(\lambda)\,d\lambda=
\sum_{k=1}^N\underset{\la=\la^k}{\rm Res}\left[\frac{
p(\lambda)^2\phi_k(\lambda)\,d\la}{(p(\la)-v^i)(p(\la)-v^j)}\right]\\
=\sum_{k=1}^N\underset{p=v^k}{\rm Res}\left[\frac{p^2\frac{\partial \la}{\partial p}}{(p-v^i)(p-v^j)}\,
\,\phi_k(\lambda(p))\,\,d p\right]=\phi_i(\la_i)\,\left(v^i\right)^2\lambda''(v^i)\delta^{ij},
\end{gather*}
the last step being due to the fact that $p=v^k$ are critical points of 
$\la$, so that the differential turns out to be regular at all these 
points for $i\neq j$, and also for $i=j$ and $k\neq i$. Making use of Lemma \ref{ndderprop}, we obtain the desired result.
\epf
\newline

Therefore the purely nonlocal operators associated to the reductions of the dispersionless 2D Toda hierarchy are
\beq
\Pi^{ij}=\frac{1}{2\pi  i}\sum_{k=1}^N\int_{C_k} \frac{p(\la)}{p(\la)-v^i} \la_{i,x}  \left(\frac{d}{d
x}\right)^{-1} \frac{p(\la)}{p(\la)-v^j} \la_{j,x}  \phi_k(\lambda)\,d\lambda.\notag
\eeq

%
\section{An example: the Dispersionless Toda chain}
The simplest example of reduction is the dispersionless Toda chain, which is given by the constraint 
$$\la = \bla = p+ v+ \frac{e^u}{p}.$$
The characteristic velocities are given by the critical points
$$v^1 =e^{\frac{u}2} ,\quad v^2 =-e^{\frac{u}2}$$
and the Riemann invariants by the critical values of $\la=\bla$
$$\la_1 = v + 2 e^{\frac{u}2}, \quad \la_2 = v- 2 e^{\frac{u}2}.$$
In this simple example we can explicitly write $\la$, $\bla$, the characteristic velocities, $u_0$ and $\varphi$ in terms of the Riemann invariants, i.e. $$\la = \bla = p + \frac{\la_1 + \la_2}2 + \left( \frac{\la_1-\la_2}4 \right)^2 p^{-1}$$
and
$$v^1=\frac{\la_1-\la_2}4 , \quad v^2=\frac{\la_2-\la_1}4, \quad 
u_0 = v = \frac{\la_1+\la_2}2, \quad
\varphi= u = 2 \log \frac{\la_1-\la_2}4.$$
It is easy to check that these functions satisfy Loewner \eq{loewner} and Gibbons-Tsarev \eq{GT} equations.

Let us compute the Hamiltonian operators associated to the metrics
$$
g_{ij}=\f{\d_i u}{\lambda_i^k} \de_{ij}
$$
for $k\geq0$, which clearly solve \eq{meq}. Explicitly
\beq
g_{11} =\frac{2 \la_1^{-k}}{\la_1-\la_2} , \quad
g_{22} =\frac{2 \la_2^{-k}}{\la_2-\la_1} .\nn
\eeq
In this case the curvature can be expressed as
\begin{equation*}
R^{12}_{12}=-\frac{v^1 v^2}{2\pi i}\,\int_C\frac{\left(\frac{\partial p}{\partial\la}\right)^2\,\la^k}
{(p(\la)-v^1)^2(p(\la)-v^2)^2}\,d\la,
\end{equation*}
or alternatively as
\begin{equation*}
R^{12}_{12}=-v^1 v^2\,\,\sum_{i=1}^2\underset{p=v^i}{\rm{Res}}\left(\frac{\lambda(p)^k\,\,\,\frac{1}{\la^{'}(p)}}{(p-v^1)^2(p-v^2)^2}\,dp\right).
\end{equation*}

For $k=0,1,2$, the abelian differential
\begin{equation*}
\frac{\lambda(p)^k\,\,\,\frac{1}{\la^{'}(p)}}{(p-v^1)^2(p-v^2)^2}\,dp
\end{equation*}
has poles only at the points  $p=v^1,\,\,$ $p=v^2,\,\,$ and therefore the curvature vanishes
 and the associated Hamiltonian operators are local.

For $k>2$ new poles appear at $p=0$ and $p=\infty$.
 Since the sum of the residues of an abelian differential on a compact Riemann surface is zero, we can substitute the sum of residues at $p=v^1,v^2$ with minus the sum of residues at $p=0$ and $p=\infty$, obtaining
\beq
R^{12}_{12}=v^1v^2\,\left(\underset{p=0}{\rm{Res}}+\underset{p=\infty}{\rm{Res}}\right)
\frac{\lambda(p)^k\,\frac{1}{\la^{'}(p)}}{(p-v^1)^2(p-v^2)^2}\,dp. \nn
\eeq
Taking into account that $\lambda(0)=\lambda(\infty)=\infty$, we easily obtain the counterpart
 of the above formulae in the $\lambda$-plane
\beq
R^{12}_{12}=2 \underset{\la=\infty}{\res} \frac{v^1 \frac{\pa p}{\pa \la} v^2 \frac{\pa p}{\pa \la} \la^k}{(p-v^1)^2(p-v^2)^2} d\la
=2\,\underset{\la=\infty}{\rm{Res}}\,\Big(w^1(\lambda)w^2(\lambda)\lambda^k\,d\lambda\Big). \nn
\eeq

Since the expansions of $w^1(\lambda)$ and $w^2(\lambda)$ near $\lambda=\infty$ have the form
\bea
&w^1(\lambda)=\sum_{k=1}^{\infty} k v^1_{(k)} \la^{-k-1}=
\f{\lambda_1-\lambda_2}{4\lambda^2}+\f{(\lambda_1-\lambda_2)(3\lambda_1+\lambda_2)}{8\lambda^3}
+\dots \nn \\
&w^2(\lambda)=\sum_{k=1}^{\infty} k v^2_{(k)} \la^{-k-1}=
\f{\lambda_2-\lambda_1}{4\lambda^2}+\f{(\lambda_2-\lambda_1)(3\lambda_2+\lambda_1)}{8\lambda^3}
+\dots \nn
\eea
we obtain the quadratic expansion of the Riemann tensor
$$
R^{12}_{12} = -2 \sum_{l+s=k-1} l v^1_{(l)} s v^2_{(s)} 
$$
and we can immediately write the corresponding nonlocal Hamiltonian operator $\Pi^{ij}$. From the last formula we have that the nonlocal tail of e.g. $\Pi^{12}$ is given by
\beq
-2 \sum_{l+s=k-1} l v^1_{(l)} \la_{1,x} \big( \frac{d}{dx} \big)^{-1} s v^2_{(s)} \la_{2,x}. \nn
\eeq
Similar formulas can be obtained in this example for purely nonlocal Hamiltonian structures.

\bigskip
{\bf Acknowledgements}
\medskip

We are grateful to John Gibbons for his interest in this work; many ideas of the present paper have been derived from a previous collaboration of two of us with him. G.~Carlet and P.~Lorenzoni wish to thank GNFM's \lq\lq Progetto Giovani\rq\rq grant, for the stay of G.Carlet at University of Milano-Bicocca. G.~Carlet acknowledges also support from the ESF-MISGAM exchange grant n.~2326 for his visit at Universidad Complutense de Madrid, where this work has been partially carried out. A.~Raimondo wishes to thank the ESF-MISGAM project for the exchange grants n.~2264 for his staying at International School for Advanced Studies, Trieste, and n.2265 for his visit at University of Milano-Bicocca.

\appendix

\end{document}